\documentclass[aps,prd,twocolumn,eqsecnum,showpacs,amsmath]{revtex4}
\usepackage[dvips]{color,graphicx}
\usepackage{amsfonts,amssymb,theorem}
\newcommand{\qed}{\hbox{\rule[-2pt]{6pt}{6pt}}}
\newcommand{\D}{{\rm d}}

{\theorembodyfont{\upshape}
\newtheorem{Prop}{Proposition}}
{\theorembodyfont{\upshape}
}
{\theorembodyfont{\upshape}
}
{\theorembodyfont{\upshape}
}
{\theorembodyfont{\upshape}
\newtheorem{Coro}{Corollary}}
{\theorembodyfont{\upshape}
}

\newcommand{\dalm}{\kern1pt\vbox{\hrule height 0.9pt\hbox{\vrule width
0.9pt\hskip 2.5pt\vbox{\vskip 5.5pt}\hskip 3pt\vrule width 0.3pt}\hrule height
0.3pt}\kern1pt}

\begin{document}

\title{
Hawking-Ellis type of matter on Killing horizons in symmetric spacetimes
}

\author{Hideki Maeda}
\email{h-maeda@hgu.jp}


\affiliation{
Department of Electronics and Information Engineering, Hokkai-Gakuen University, Sapporo 062-8605, Japan
}

\date{\today}

\begin{abstract}
Spherically, plane, or hyperbolically symmetric spacetimes with an additional hypersurface orthogonal Killing vector are often called ``static'' spacetimes even if they contain regions where the Killing vector is non-timelike.
It seems to be widely believed that an energy-momentum tenor for a matter field compatible with these spacetimes in general relativity is of the Hawking-Ellis type I everywhere.
We show in arbitrary $n(\ge 3)$ dimensions that, contrary to popular belief, a matter field on a Killing horizon is not necessarily of type I but can be of type II.
Such a type-II matter field on a Killing horizon is realized in the Gibbons-Maeda-Garfinkle-Horowitz-Strominger black hole in the Einstein-Maxwell-dilaton system and may be interpreted as a mixture of a particular anisotropic fluid and a null dust fluid.
\end{abstract}


\maketitle


\section{Introduction}
Spherically symmetric spacetime has been one of the most important classes in the research of general relativity because it is simple enough but shows us a variety of non-trivial properties of spacetime.
(See textbooks~\cite{Stephani:2003,GriffithsPodolsky:2009} for example.)
In fact, the Schwarzschild vacuum spacetime is the simplest model of an asymptotically flat black hole and its maximal extension exposed the nature of the event horizon and the central singularity.
In addition, its charged version, the Reissner-Nordstr\"om spacetime, exposed the existence of a naked singularity surrounded by an additional inner horizon~\cite{GriffithsPodolsky:2009}.
The generalized Schwarzschild spacetime with planar or hyperbolic symmetry instead of spherical symmetry does not describe a black hole but a naked singularity.
However, in the presence of a negative cosmological constant, the spacetime can represent an asymptotically anti-de~Sitter black hole~\cite{Lemos:1994xp,Huang:1995zb,Vanzo:1997gw}.
Such black holes with non-spherical symmetry are called topological black holes.

All these spacetimes admit a hypersurface orthogonal Killing vector $\xi^\mu$ in addition to a set of Killing vectors which generates a maximal spatial symmetry.
The event horizon of a black hole in these spacetimes is a Killing horizon where $\xi^\mu$ becomes null.
If the Killing horizon is non-degenerate, the spacetime contains a dynamical region where $\xi^\mu$ becomes spacelike.
Even in such cases, these spacetimes are often called ``static'' spacetimes.

This class of $n(\ge 3)$-dimensional ``static'' spacetimes can be represented generally in the following Buchdahl coordinates:
\begin{align}
\label{metric-Buchdahl}
\D s^2=&-H(x)\D t^2+\frac{\D x^2}{H(x)}+r(x)^2\gamma_{ij}(z)\D z^i\D z^j,
\end{align} 
where $\gamma_{ij}(z)$ ($i,j=2,3,\cdots,n-1$) is the metric on a $(n-2)$-dimensional maximally symmetric space $K^{n-2}$.
Throughout this paper, we assume $r\ge 0$ without loss of generality.
The Riemann tensor of $K^{n-2}$ is given by 
\begin{align}
{}^{(n-2)}R^{ij}_{~~kl}=k(\delta^i_{~k}\delta^j_{~l}-\delta^i_{~l}\delta^j_{~k}),
\end{align} 
where $k$ takes $1$, $0$, and $-1$ corresponding to spherical, planar, and hyperbolic symmetry, respectively.
The spacetime (\ref{metric-Buchdahl}) admits a hypersurface orthogonal Killing vector $\xi^\mu(\partial/\partial x^\mu)=\partial/\partial t$, of which squared norm is given by $\xi_\mu\xi^\mu=-H(x)$.
Hence, $\xi^\mu$ is timelike (spacelike) in an untrapped (trapped) region defined by $H(x)>(<)0$.
A Killing horizon associated with $\xi^\mu$ is a regular null hypersurface $x=x_{\rm h}$ satisfying $H(x_{\rm h})=0$.

Now let us consider an energy-momentum tenor ${T}_{\mu\nu}$ of a matter field compatible with the spacetime (\ref{metric-Buchdahl}) in general relativity.
This is equivalent to consider an effective energy-momentum tensor defined by ${T}_{\mu\nu}:=G_{\mu\nu}$ (with units such that $c=8\pi G=1$) in generalized theories of gravity.
In general, an energy-momentum tensor $T_{\mu\nu}$ can be classified into four types depending on the properties of its eigenvectors in arbitrary $n(\ge 3)$ dimensions~\cite{Hawking:1973uf,Santos:1995kt,Reboucas:2003fx}. 
All the four types of $T_{\mu\nu}$ in this Hawking-Ellis classification are summarized in Table~\ref{table:scalar+1}. 
\begin{table}[htb]
\begin{center}
\caption{\label{table:scalar+1} Eigenvectors of type-I--IV energy-momentum tensors and their expressions in the Segre notation. (See appendix in~\cite{Maeda:2018hqu} for details.)}
\scalebox{0.90}{
\begin{tabular}{|c|c|c|c|}
\hline
Type & Eigenvectors & Segre notation \\ \hline\hline
I & 1 timelike, $n-1$ spacelike & $[1,11\cdots1]$ \\ \hline
II & 1 null (doubly degenerated), $n-2$ spacelike & $[211\cdots1]$ \\ \hline
III & 1 null (triply degenerated), $n-3$ spacelike & $[311\cdots 1]$ \\ \hline
IV & 2 complex, $n-2$ spacelike & $[Z{\bar Z}1\cdots 1]$ \\ 
\hline
\end{tabular} 
}\end{center}
\end{table} 

Among these four types, orthonormal components $T^{(a)(b)}=T^{\mu\nu}E_\mu^{(a)}E_\nu^{(b)}$ of the Hawking-Ellis type-I energy-momentum tensor in the local Lorentz frame can be written in the following canonical form~\cite{Hawking:1973uf,Maeda:2018hqu}:
\begin{equation} 
\label{T-typeI}
T^{(a)(b)}=\mbox{diag}(\rho,p_1,p_2,\cdots,p_{n-1}).
\end{equation}
Here ${E}_\mu^{(a)}~(a=0,1,\cdots,n-1)$ are orthonormal basis one-forms satisfying ${E}^\mu_{(a)}{E}_{(b)\mu}=\eta_{(a)(b)}$, where $\eta_{(a)(b)}$ is the Minkowski metric in the local Lorentz frame and the spacetime metric $g_{\mu\nu}$ is given by $g_{\mu\nu}=\eta_{(a)(b)}E^{(a)}_{\mu}E^{(b)}_{\nu}$.
On the other hand, the canonical form of orthonormal components of the type II energy-momentum tensor is 
\begin{equation} 
\label{T-typeII}
T^{(a)(b)}=\left( 
\vphantom{\begin{array}{c}1\\1\\1\\1\\1\\1\end{array}}
\begin{array}{cccccc}
\rho+\nu &\nu&0&0&\cdots &0\\
\nu&-\rho+\nu&0&0&\cdots &0\\
0&0&p_2&0&\cdots&0 \\
0&0&0&\ddots&\vdots&\vdots \\
\vdots&\vdots&\vdots&\cdots&\ddots&0\\
0&0&0 &\cdots&0&p_{n-1}
\end{array}
\right).
\end{equation}
Note that a type II energy-momentum tensor reduces to type I (\ref{T-typeI}) with $p_1=-\rho$ if $\nu=0$.

In this context, it seems to be widely believed that ${T}_{\mu\nu}(=:G_{\mu\nu})$ in the spacetime (\ref{metric-Buchdahl}) is of the Hawking-Ellis type I {\it everywhere}.
For example, this claim has been made in~\cite{Visser:1992qh} for $n=4$ with $k=1$.
It is certainly true in a region with $H(x)\ne 0$.
However, as shown in the present paper, it is not the case on a Killing horizon $x=x_{\rm h}$.
The incorrect conclusion in~\cite{Visser:1992qh} stems from the analysis based on a singular coordinate system at $x=x_{\rm h}$ like the coordinates (\ref{metric-Buchdahl}).
In fact, taking a limit $x\to x_{\rm h}$ in such a singular coordinate system does not always lead to a correct result at $x=x_{\rm h}$.

This situation is similar to the computation to derive the surface gravity $\kappa$ on a Killing horizon, defined by $\xi^\nu\nabla_\nu \xi^\mu|_{x=x_{\rm h}}=\kappa\xi^\mu|_{x=x_{\rm h}}$.
In the coordinate system (\ref{metric-Buchdahl}) with $\xi^\mu(\partial/\partial x^\mu)=\partial/\partial t$, we obtain $\xi^\nu\nabla_\nu \xi^t=0=\xi^\nu\nabla_\nu \xi^i$ and $\xi^\nu\nabla_\nu \xi^x=HH'/2$, where a prime denotes differentiation with respect to $x$.
Hence, the singular coordinate system (\ref{metric-Buchdahl}) leads to a wrong conclusion $\kappa=0$.
Of course, adopting a regular coordinate system at $x=x_{\rm h}$ with advanced time $v$ or retarded time $u$, we obtain the correct result $\kappa=H'(x_{\rm h})/2$ with $\xi^\mu(\partial/\partial x^\mu)=\partial/\partial v$ or $\kappa=-H'(x_{\rm h})/2$ with $\xi^\mu(\partial/\partial x^\mu)=\partial/\partial u$.
Advantages of the Buchdahl coordinates (\ref{metric-Buchdahl}) as the ``quasiglobal'' coordinates have been emphasized in~\cite{Bronnikov:1998gf,Bronnikov:1998hm,Bronnikov:2001tv,Bronnikov:2008ia}.

In the present paper, we will prove some of generic properties of a matter field in the spacetime (\ref{metric-Buchdahl}) on and off a Killing horizon.
In Sec.~\ref{sec:HE-matter}, we will derive a necessary and sufficient condition for that ${T}_{\mu\nu}$ is of the Hawking-Ellis type II on a Killing horizon.
In Sec.~\ref{sec:EC}, we will discuss the standard energy conditions on and off a Killing horizon.
A result in Sec.~\ref{sec:HE-matter} will be applied to ``static'' perfect-fluid solutions in Sec.~\ref{sec:PF}.
We will summarize our results in the final section.
Our conventions for curvature tensors are $[\nabla _\rho ,\nabla_\sigma]V^\mu ={R^\mu }_{\nu\rho\sigma}V^\nu$ and $R_{\mu \nu }={R^\rho }_{\mu \rho \nu }$.
The signature of the Minkowski spacetime is $(-,+,\ldots,+)$, and Greek indices run over all spacetime indices.

We note that all the following results remain valid even if the base manifold $K^{n-2}$ in the spacetime (\ref{metric-Buchdahl}) is replaced by an arbitrary Einstein space, of which Ricci tensor satisfies ${}^{(n-2)}R_{ij}=k(n-3)\gamma_{ij}$, due to the fact that the Riemann tensor ${R^\mu}_{\nu\rho\sigma}$ does not appear explicitly in the definition of the Einstein tensor $G_{\mu\nu}$.

\section{Main results}
\subsection{Hawking-Ellis types of matter}
\label{sec:HE-matter}

First we prove that the matter field is of type I in a region with $H(x)\ne 0$ in the spacetime (\ref{metric-Buchdahl}).

\begin{Prop}
\label{Prop:matter-offhorizon}
In a region with $H(x)\ne 0$ in the spacetime (\ref{metric-Buchdahl}), the energy-momentum tensor ${T}_{\mu\nu}(:=G_{\mu\nu})$ is of the Hawking-Ellis type I (\ref{T-typeI}) with $p_2=p_3=\cdots=p_{n-1}$.
\end{Prop}
{\it Proof:}
In a region with $H(x)\ne 0$, we introduce the following orthonormal basis one-forms in the local Lorentz frame satisfying ${E}_\mu^{(a)}{E}^{(b)\mu}=\mbox{diag}(-1,1,\cdots,1)$:
\begin{align}
&E^{(0)}_\mu\D x^\mu = \left\{
\begin{array}{ll}
-\sqrt{H}\D t & (\mbox{if}~H(x)>0)\\
-\sqrt{-H^{-1}}\D x & (\mbox{if}~H(x)<0)
\end{array}
\right.,\label{E0-1}\\
&E^{(1)}_\mu\D x^\mu = \left\{
\begin{array}{ll}
-\sqrt{H^{-1}}\D x & (\mbox{if}~H(x)>0)\\
-\sqrt{-H}\D t & (\mbox{if}~H(x)<0)
\end{array}
\right.,\label{E1-1}\\
&{E}_\mu^{(k)}\D x^\mu=r{e}_i^{(k)}\D z^i, \label{E23-1}
\end{align}
where ${e}_i^{(k)}~(k=2,3,\cdots,n-1)$ are basis one-forms on $K^{n-2}$ satisfying 
\begin{align}
\gamma_{ij}=\delta_{(k)(l)}e^{(k)}_{i}e^{(l)}_{j}~\leftrightarrow~\gamma^{ij}e^{(k)}_{i}e^{(l)}_{j}=\delta^{(k)(l)}.\label{e-K}
\end{align}
Non-zero components of the Einstein tensor of the spacetime (\ref{metric-Buchdahl}) are given by 
\begin{align}
G^t_{~t}=&\frac{n-2}{2r^2}\biggl[rr'H' +2rr''H+(n-3)(H{r'}^2-k)\biggl],\\
G^x_{~x}=&\frac{n-2}{2r^2}\left[rr'H'+(n-3)(H{r'}^2-k)\right],\\
G^i_{~j}=&\delta^i_{~j}p_{\rm t}(x),
\end{align}
where $p_{\rm t}(x)$ is defined by 
\begin{align}
p_{\rm t}(x):=&\frac12r^{-2}\biggl[r^2H''+2(n-3)rr'H' \nonumber \\
&+2(n-3)rr''H+ (n-3)(n-4)(H{r'}^2-k)\biggl].\label{def-pt}
\end{align}
Then, $T^{(a)(b)}(=G^{\mu\nu}E^{(a)}_{\mu}E^{(b)}_{\nu})$ is given in the type-I form~(\ref{T-typeI}) with
\begin{align}
\rho =& -\frac{n-2}{2r^2}\biggl[rr'H' +(H+|H|)rr'' \nonumber \\
&~~~~~~~~~~~~~~~~~+(n-3)(H{r'}^2-k)\biggl]=:\eta(x),\label{bounce-rho}\\
p_1 =& \frac{n-2}{2r^2}\biggl[rr'H'+(H-|H|)rr'' \nonumber \\
&~~~~~~~~~~~~~~~~~+(n-3)(H{r'}^2-k)\biggl]=:p_{\rm r}(x),\label{bounce-p1}\\
&p_2=p_3=\cdots=p_{n-1}=p_{\rm t}(x).\label{bounce-p2}
\end{align}
\qed
\bigskip

Such a type-I matter field in Proposition~\ref{Prop:matter-offhorizon} can be interpreted as an anisotropic fluid, of which energy-momentum tensor is given by 
\begin{align}
{T}_{\mu\nu}=&(\rho+p_2)u_\mu u_\nu+(p_1-p_2)s_\mu s_\nu +p_2g_{\mu\nu},\label{Tab-a}
\end{align}
where $u_\mu u^\mu=-1$, $s_\mu s^\mu=1$, and $u_\mu s^\mu=0$.
Adopting orthonormal basis one-forms $E_\mu^{(0)}$ and $E_\mu^{(1)}$ such that
\begin{align}
u_\mu={E}_\mu^{(0)}, \quad s_\mu={E}_\mu^{(1)},\label{us-E}
\end{align}
we obtain $T^{(a)(b)}=T^{\mu\nu}E^{(a)}_{\mu}E^{(b)}_{\nu}$ in the type-I form~(\ref{T-typeI}) with $p_2=p_3=\cdots=p_{n-1}$.

It should be emphasized that Proposition~\ref{Prop:matter-offhorizon} cannot be directly applied to a Killing horizon defined by $H(x_{\rm h})=0$.
This is because $x=x_{\rm h}$ is a coordinate singularity in the coordinate system (\ref{metric-Buchdahl}).
Indeed, Eqs.~(\ref{E0-1}) and (\ref{E1-1}) show that either of $E^{(0)}_\mu$ or $E^{(1)}_\mu$ diverges there.

Regular coordinate systems covering a Killing horizon are obtained by introducing advanced time $v$ or retarded time $u$ defined by 
\begin{align}
&v:=t+\int H(x)^{-1}\D x,\label{def-v}\\
&u:=t-\int H(x)^{-1}\D x.\label{def-u}
\end{align} 
Then, the spacetime (\ref{metric-Buchdahl}) is written as
\begin{align}
\D s^2=-H(x)\D v^2+2\D v\D x+r(x)^2\gamma_{ij}(z)\D z^i\D z^j,\label{metric-Buchdahl-v}\\
\D s^2=-H(x)\D u^2-2\D u\D x+r(x)^2\gamma_{ij}(z)\D z^i\D z^j,\label{metric-Buchdahl-u}
\end{align} 
in which the metric and its inverse are both finite at $x=x_{\rm h}$.
Now we prove that the matter field on a Killing horizon can be of type II as well.


\begin{Prop}
\label{Prop:matter-on-horizon}
On a Killing horizon defined by $H(x_{\rm h})=0$ in the spacetime (\ref{metric-Buchdahl}), the energy-momentum tensor ${T}_{\mu\nu}(:=G_{\mu\nu})$ is of the Hawking-Ellis type I (\ref{T-typeI}) with $p_1= -\rho$ and $p_2=p_3=\cdots=p_{n-1}$ if $r''(x_{\rm h})=0$ holds.
If $r''(x_{\rm h})\ne 0$ holds, it is of type II (\ref{T-typeII}) with $p_2=p_3=\cdots=p_{n-1}$.
\end{Prop}
{\it Proof:}
We introduce the following orthonormal basis one-forms in the local Lorentz frame satisfying ${E}_\mu^{(a)}{E}^{(b)\mu}=\mbox{diag}(-1,1,\cdots,1)$ in the coordinate system (\ref{metric-Buchdahl-v}):
\begin{align}
{E}_\mu^{(0)}\D x^\mu=&-\frac{1}{\sqrt{2}}\left(1+\frac{H}{2}\right)\D v+\frac{1}{\sqrt{2}}\D x,\label{basis0-H}\\
{E}_\mu^{(1)}\D x^\mu=&-\frac{1}{\sqrt{2}}\left(1-\frac{H}{2}\right)\D v-\frac{1}{\sqrt{2}}\D x,\\
{E}_\mu^{(k)}\D x^\mu=&r{e}_i^{(k)}\D z^i,\label{basis2-H}
\end{align}
where basis one-forms ${e}_i^{(k)}~(k=2,3,\cdots,n-1)$ on $K^{n-2}$ satisfy Eq.~(\ref{e-K}).
Non-zero components of the Einstein tensor of the spacetime (\ref{metric-Buchdahl-v}) are given by 
\begin{align}
G^{vv}=&-(n-2)r^{-1}r'',\label{Gvv}\\
G^{vx}=&G^{xv}=\frac{n-2}{2r^2}\left[rr'H'+(n-3)(H{r'}^2-k)\right],\\
G^{xx}=&\frac{(n-2)H}{2r^2}\left[rr'H'+(n-3)(H{r'}^2-k)\right],\\
G^{ij}=&\gamma^{ij}r^{-2}p_{\rm t}(x),
\end{align}
where $p_{\rm t}(x)$ is defined by Eq.~(\ref{def-pt}).
Then, non-zero components of $T^{(a)(b)}(=G^{\mu\nu}E^{(a)}_{\mu}E^{(b)}_{\nu})$ are computed to give
\begin{align}
T^{(0)(0)}=&-\frac{n-2}{8r^2}\{4rr'H'+rr''(H+2)^2 \nonumber \\
&~~~~~~+4(n-3)(H{r'}^2-k)\},\label{T00-h}\\
T^{(0)(1)}=&T^{(1)(0)}=\frac{n-2}{8r}r''(H^2-4),\\
T^{(1)(1)}=&\frac{n-2}{8r^2}\{4rr'H'-rr''(H-2)^2 \nonumber \\
&~~~~~~+4(n-3)(H{r'}^2-k)\},\\
T^{(i)(j)}=&\delta^{(i)(j)}p_{\rm t}(x).\label{Tij-h}
\end{align}
Equations~(\ref{T00-h})--(\ref{Tij-h}) show that $T^{(a)(b)}$ is in the type-II form~(\ref{T-typeII}) on a Killing horizon $H(x_{\rm h})=0$ with
\begin{align}
&\rho=\eta(x_{\rm h})(=-p_{\rm r}(x_{\rm h})),\label{Tab-horizon-rho}\\
&\nu=-\frac{n-2}{2}r^{-1}r''|_{x=x_{\rm h}},\label{Tab-horizon-nu}\\
&p_2=p_3=\cdots=p_{n-1}=p_{\rm t}(x_{\rm h}).\label{Tab-horizon-p2}
\end{align}
where $\eta(x)$ and $p_{\rm r}(x)$ are defined by Eqs.~(\ref{bounce-rho}) and (\ref{bounce-p1}), respectively.
Therefore, ${T}_{\mu\nu}$ is of type I if $r''(x_{\rm h})=0$ and of type II otherwise.
The same result is obtained in the coordinate system (\ref{metric-Buchdahl-u}) by a coordinate transformation $v=-u$.
\qed
\bigskip

Such a type-II matter field on a Killing horizon in Proposition~\ref{Prop:matter-on-horizon} can be interpreted as a mixture of an anisotropic fluid (\ref{Tab-a}) with $p_1=-\rho$ and a null dust fluid, of which energy-momentum tensor is given by 
\begin{align}
{T}_{\mu\nu}|_{x=x_{\rm h}}=&(\rho+p_2)(u_\mu u_\nu-s_\mu s_\nu) +p_2g_{\mu\nu} \nonumber \\
&+\mu k_\mu k_\nu,\label{Tab-1}
\end{align}
where $k_\mu k^\mu=0$.
The last term in Eq.~(\ref{Tab-1}) is the energy-momentum tensor of a null dust fluid with its energy density $\mu$.
In terms of orthonormal basis one-forms $E_\mu^{(0)}$ and $E_\mu^{(1)}$ satisfying Eq.~(\ref{us-E}), we represent $k_\mu$ as
\begin{align}
k_\mu=\frac{1}{\sqrt{2}}({E}_\mu^{(0)}-{E}_\mu^{(1)}).\label{kl-E}
\end{align}
Then orthonormal components $T^{(a)(b)}|_{x=x_{\rm h}}$ are obtained in the form of Eq.~(\ref{T-typeII}) with $\nu=\mu/2$ and $p_2=p_3=\cdots=p_{n-1}$.
Using the basis one-forms (\ref{basis0-H})--(\ref{basis2-H}) at $x=x_{\rm h}$ in the spacetime (\ref{metric-Buchdahl-v}), we obtain
\begin{align}
&k_\mu\D x^\mu=\D x,\qquad k^\mu\frac{\partial}{\partial x^\mu}=\frac{\partial}{\partial v},
\end{align}
so that a null dust fluid is confined on a Killing horizon.
$\rho$ and $p_2$ in Eq.~(\ref{Tab-1}) are given by Eqs.~(\ref{Tab-horizon-rho}) and (\ref{Tab-horizon-p2}), respectively, while Eq.~(\ref{Tab-horizon-nu}) gives $\mu$ as
\begin{align}
\mu=&2\nu=-(n-2)r^{-1}r''|_{x=x_{\rm h}}.\label{mu-horizon}
\end{align}

Of course, the spacetime (\ref{metric-Buchdahl}) with a Killing horizon can be a solution with a different matter field.
It is well known that the Reissner-Nordstr\"om black hole in the Einstein-Maxwell system and its higher-dimensional and topological generalization are described by the metric (\ref{metric-Buchdahl}) with $r(x)=x$.
By Proposition~\ref{Prop:matter-on-horizon}, the Maxwell field is of type I on a Killing horizon in those spacetimes.
This is also the case with the charged BTZ black hole in three dimensions ($n=3$)~\cite{Banados:1992wn,Martinez:1999qi}.

In contrast, the matter field is of type II on a Killing horizon of the Gibbons-Maeda-Garfinkle-Horowitz-Strominger (GM-GHS) dilatonic black hole~\cite{Gibbons:1987ps,Garfinkle:1990qj} in the following four-dimensional ($n=4$) Einstein-Maxwell-dilaton system;
\begin{align} 
S&=\int \D^4x\sqrt{-g}\biggl[R-2(\nabla \phi)^2-e^{-2\alpha\phi}F_{\mu\nu}F^{\mu\nu}\biggl],\label{action}
\end{align}
where $\alpha$ is the dilaton coupling constant and $F_{\mu\nu}=\partial_\mu A_\nu-\partial_\nu A_\mu$.
The GM-GHS solution is given by 
\begin{align}
&\D s^2=-\frac{f}{B}\D t^2+\frac{B}{f}\D x^2+Bx^2\D\Omega^2, \label{gm1} \\
&A_\mu\D x^\mu=\frac{2m \chi}{(1-\chi^2)\sqrt{1+\alpha^2}}\frac{B^{-(1+\alpha^2)/2}}{x}\D t,\\
&\phi=-\frac{\alpha}{1+\alpha^2}\ln \biggl(1+\frac{\chi^2}{1-\chi^2}\frac{2m}{x}\biggl),\label{phi-GM}
\end{align} 
where $\D\Omega^2:=\D\theta^2+\sin^2\theta\D\varphi^2$, $m$ and $\chi$ are constants, and the metric functions $f(x)$ and $B(x)$ are given by 
\begin{align}
&f(x):=1-\frac{2m}{x},\label{f-func}\\
&B(x):=\biggl(1+\frac{\chi^2}{1-\chi^2}\frac{2m}{x}\biggl)^{2/(1+\alpha^2)}.\label{B-func}
\end{align} 

Equations~(\ref{gm1})--(\ref{B-func}) solve the following field equations in the system (\ref{action});
\begin{align}
&{G}_{\mu\nu}=T_{\mu\nu},\quad \nabla_\nu(e^{-2\alpha\phi}F^{\mu\nu})=0,\label{beq}\\
&\dalm \phi+\frac{\alpha}{2}e^{-2\alpha\phi}F_{\rho\sigma}F^{\rho\sigma}=0 \label{phi}
\end{align}
with the energy-momentum tensor $T_{\mu\nu}$ given by 
\begin{align}
T_{\mu\nu}=&2\nabla_\mu \phi \nabla_\nu\phi-g_{\mu\nu}(\nabla\phi)^2 \nonumber \\
&+2e^{-2\alpha\phi}\biggl(F_{\mu\rho}F_{\nu}^{~\rho}-\frac14 g_{\mu\nu}F_{\rho\sigma}F^{\rho\sigma}\biggl). \label{T-dilaton}
\end{align}
We note that, by a coordinate transformation $x=2m/(1-e^{2my})$, the untrapped region ($f>0$) of the GM-GHS solution (\ref{gm1})--(\ref{B-func}) is transformed into a subclass (with $\kappa>0$) of the solution obtained by Bronnikov and Shikin in 1977~\cite{Bronnikov:1977is}. 
(See~\cite{Bronnikov:1977is-English} for a translated version in English.)
Thus, the GM-GHS solution is an analytic extension of the Bronnikov-Shikin solution with $\kappa>0$ into the trapped region ($f<0$).

For $m>0$ and $0<\chi^2<1$, the GM-GHS spacetime (\ref{gm1}) represents an asymptotically flat black hole with a single Killing horizon at $x=2m(>0)$.
Since the areal radius $r(x)=xB(x)^{1/2}$ gives
\begin{align}
r''(2m)=-\frac{\chi^4\alpha^2}{2m(1+\alpha^2)^2}\biggl(\frac{1}{1-\chi^2}\biggl)^{1/(1+\alpha^2)},\label{r''-dilaton}
\end{align} 
the energy-momentum tensor (\ref{T-dilaton}) on the Killing horizon is of type II for $\alpha\ne 0$ by Proposition~\ref{Prop:matter-on-horizon}.
For $\alpha=0$, the GM-GHS solution becomes the Reissner-Nordstr\"om solution with a trivial dilaton field $\phi\equiv 0$ and then the energy-momentum tensor (\ref{T-dilaton}) is of type I on a Killing horizon.

In the GM-GHS solution, $\nabla_\mu\phi$ becomes null at the Killing horizon $x=2m$ and $G^{vv}=T^{vv}$ with Eq.~(\ref{Gvv}) shows $-r^{-1}r''={\phi'}^2$.
Then, it is confirmed by Eq.~(\ref{Tab-horizon-nu}) that a non-trivial dilaton field $\phi$ makes $T_{\mu\nu}$ type II on the Killing horizon.

\subsection{Energy conditions}
\label{sec:EC}

Propositions~\ref{Prop:matter-offhorizon} and \ref{Prop:matter-on-horizon} assert that, as realized in the GM-GHS black hole (\ref{gm1}), a matter field on a Killing horizon in the spacetime (\ref{metric-Buchdahl}) can be of the Hawking-Ellis type II, while a matter off the horizon is always of type I.
Such black holes may be realized also in generalized theories of gravity.
In such a case, one can define an effective energy-momentum tensor by ${T}_{\mu\nu}:={G}_{\mu\nu}$ and check the standard energy conditions for ${T}_{\mu\nu}$ as a measure of deviation from general relativity.
In this section, we discuss the energy conditions for ${T}_{\mu\nu}(:={G}_{\mu\nu})$ in the spacetime (\ref{metric-Buchdahl}).

The standard energy conditions consist of the {\it null} energy condition (NEC), {\it weak} energy condition (WEC), {\it dominant} energy condition (DEC), and {\it strong} energy condition (SEC).
Equivalent expressions of these conditions for the type-I energy-momentum tensor (\ref{T-typeI}) are given by
\begin{align}
\mbox{NEC}:&~~\rho+p_i\ge 0,\label{NEC-I}\\
\mbox{WEC}:&~~\rho\ge 0\mbox{~in addition to NEC},\label{WEC-I}\\
\mbox{DEC}:&~~\rho-p_i\ge 0\mbox{~in addition to WEC},\label{DEC-I}\\
\mbox{SEC}:&~~(n-3)\rho+\mbox{$\sum_{j=1}^{n-1}$}p_j\ge 0 \nonumber \\
&~~\mbox{~in addition to NEC}\label{SEC-I}
\end{align}
for all $i=1,2,\cdots,n-1$~\cite{Maeda:2018hqu}.
Those for the type-II energy-momentum tensor (\ref{T-typeII}) are 
\begin{align}
\mbox{NEC}:&~~\nu\ge 0\mbox{~and~}\rho+p_i\ge 0,\label{NEC-II}\\
\mbox{WEC}:&~~\rho\ge 0\mbox{~in addition to NEC},\label{WEC-II}\\
\mbox{DEC}:&~~\rho-p_i\ge 0\mbox{~in addition to WEC},\label{DEC-II}\\
\mbox{SEC}:&~~(n-4)\rho+\mbox{$\sum_{j=2}^{n-1}$}p_j\ge 0 \nonumber \\
&~~\mbox{~in addition to NEC}\label{SEC-II}
\end{align}
for all $i=2,3,\cdots,n-1$~\cite{Maeda:2018hqu}.

The following proposition shows that inequalities derived from Eqs.~(\ref{NEC-I})--(\ref{SEC-I}) for the spacetime (\ref{metric-Buchdahl}) in the region with $H(x)\ne 0$ can be conveniently used just by taking the limit $x\to x_{\rm h}$ to check the energy conditions on a Killing horizon $H(x_{\rm h})= 0$.

\begin{Prop}
\label{prop:I-II}
In the spacetime (\ref{metric-Buchdahl}), the energy conditions (\ref{NEC-II})--(\ref{SEC-II}) on a Killing horizon $H(x_{\rm h})=0$ can be obtained in the limit of $x\to x_{\rm h}$ from the energy conditions (\ref{NEC-I})--(\ref{SEC-I}) off the Killing horizon.
\end{Prop}
{\it Proof:}
Equations~(\ref{bounce-rho}) and (\ref{bounce-p1}) give
\begin{align}
&\rho+p_1 = -(n-2)r^{-1}r''|H|.\label{key-ec1}
\end{align}
By Eqs.~(\ref{key-ec1}) and (\ref{Tab-horizon-nu}), the condition $\rho+p_1\ge 0$ in the region with $H\ne 0$ and the condition $\nu\ge 0$ on a Killing horizon $H(x_{\rm h})= 0$ give the same inequality $r''\le 0$.
(Note that we have assumed $r\ge 0$ without loss of generality.)

By Eqs.~(\ref{bounce-rho})--(\ref{bounce-p2}) and (\ref{key-ec1}), the energy conditions (\ref{NEC-I})--(\ref{SEC-I}) in the region with $H\ne 0$ are equivalent to
\begin{align}
\mbox{NEC}:&~~r''(x)\le 0~\mbox{and}~\eta(x)+p_{\rm t}(x)\ge 0,\label{NEC-I-2}\\
\mbox{WEC}:&~~\eta(x)\ge 0\mbox{~in addition to NEC},\label{WEC-I-2}\\
\mbox{DEC}:&~~\eta(x)-p_{\rm t}(x)\ge 0~\mbox{~in addition to WEC},\label{DEC-I-2}\\
\mbox{SEC}:&~~(n-4)\eta(x)+(n-2)(p_{\rm t}(x)-r^{-1}r''|H|)\ge 0 \nonumber \\
&~~\mbox{~in addition to NEC}.\label{SEC-I-2}
\end{align}
By Eqs.~(\ref{Tab-horizon-rho})--(\ref{Tab-horizon-p2}), the energy conditions (\ref{NEC-II})--(\ref{SEC-II}) on a Killing horizon $H(x_{\rm h})= 0$ are equivalent to
\begin{align}
\mbox{NEC}:&~~r''(x_{\rm h})\le 0\mbox{~and~}\eta(x_{\rm h})+p_{\rm t}(x_{\rm h})\ge 0,\label{NEC-II-2}\\
\mbox{WEC}:&~~\eta(x_{\rm h})\ge 0\mbox{~in addition to NEC},\label{WEC-II-2}\\
\mbox{DEC}:&~~\eta(x_{\rm h})-p_{\rm t}(x_{\rm h})\ge 0\mbox{~in addition to WEC},\label{DEC-II-2}\\
\mbox{SEC}:&~~(n-4)\eta(x_{\rm h})+(n-2)p_{\rm t}(x_{\rm h})\ge 0 \nonumber \\
&~~\mbox{~in addition to NEC}.\label{SEC-II-2}
\end{align}
Inequalities (\ref{NEC-I-2})--(\ref{SEC-I-2}) reduce to Eqs.~(\ref{NEC-II-2})--(\ref{SEC-II-2}) in the limit to a Killing horizon $x\to x_{\rm h}$.
\qed
\bigskip

As observed in Eqs.~(\ref{NEC-I})--(\ref{SEC-II}), violation of the NEC means violation of all the standard energy conditions since NEC is the weakest one among them~\cite{Hawking:1973uf,Maeda:2018hqu}.
In the spacetime (\ref{metric-Buchdahl}), the following simple sufficient condition for the NEC violation is available, which is a generalization of the theorem in~\cite{Lobo:2020ffi} for the spacetime (\ref{metric-Buchdahl}) with $n=4$ and $k=1$ under an assumption $H(x)\ne 0$.
\begin{Prop}
\label{Prop:no-go-bounce}
In the spacetime (\ref{metric-Buchdahl}) including Killing horizons $H(x_{\rm h})=0$, all the stanadard energy conditions are violated in a region where $r''>0$ holds.
\end{Prop}
{\it Proof:}
By Eqs.~(\ref{NEC-I-2}) and (\ref{NEC-II-2}).
\qed
\bigskip

By Proposition~\ref{Prop:no-go-bounce}, $r''\le 0$ holds in a region of spacetime (\ref{metric-Buchdahl}) with a physically reasonable matter field in general relativity.
This is achieved everywhere in the case of $r(x)=x$.
This inequality also holds everywhere in the GM-GHS solution.
The metric (\ref{gm1}) gives
\begin{align}
r''(x)=-\frac{4m^2\chi^4\alpha^2}{(1+\alpha^2)^2(1-\chi^2)^2x^3}B(x)^{-(1+2\alpha^2)/2},
\end{align}
which satisfies $r''\le 0$ with equality holding for $\alpha=0$.
This is consistent with the fact that the energy-momentum tensor (\ref{T-dilaton}) in the Einstein-Maxwell-dilaton system (\ref{action}) satisfies all the standard energy conditions in the most general setting~\cite{Maeda:2018hqu}.

In contrast, by Proposition~\ref{Prop:no-go-bounce}, all the standard energy conditions are violated everywhere in the case of $r(x)=\sqrt{x^2+l^2}$ with a non-zero constant $l$, which gives $r''(x)=l^2/(r^2+l^2)^{3/2}(>0)$.
This is the case of the simplest Ellis-Bronnikov wormhole solution with a minimally coupled massless ghost scalar field, of which metric is given by Eq.~(\ref{metric-Buchdahl}) with $H(x)=1$ and $r(x)=\sqrt{x^2+l^2}$ for $n=4$ and $k=1$~\cite{Ellis:1973yv,Bronnikov:1973fh}.
Indeed, such a ghost scalar field violates all the standard energy conditions in the most general setting~\cite{Maeda:2018hqu}.

The metric ansatz (\ref{metric-Buchdahl}) with $r(x)=\sqrt{x^2+l^2}$ has also been adopted to construct a non-singular black hole of the black-bounce type~\cite{Simpson:2018tsi,Lobo:2020ffi,Fran:2021pyi}.
Proposition~\ref{Prop:no-go-bounce} shows that the effective energy-momentum tensor ${\bar T}_{\mu\nu}:=G_{\mu\nu}$ in such a spacetime violates all the standard energy conditions, as clearly stated in~\cite{Lobo:2020ffi} for the spacetime (\ref{metric-Buchdahl}) with $n=4$ and $k=1$ except on Killing horizons.

\subsection{Application to perfect-fluid solutions}
\label{sec:PF}

Proposition~\ref{Prop:matter-on-horizon} claims that a matter field on a Killing horizon can be interpreted as a mixture of a particular anisotropic fluid and a null dust fluid, of which energy-momentum tensor is given by Eq.~(\ref{Tab-1}), where $\rho$, $p_2$, and $\nu$ are given by Eqs.~(\ref{Tab-horizon-rho}), (\ref{Tab-horizon-p2}), and (\ref{mu-horizon}), respectively.
The following corollary of Proposition~\ref{Prop:matter-on-horizon} exposes a generic property of ``static'' perfect-fluid solutions that admit a Killing horizon.
\begin{Coro}
\label{Coro:p-fluid}
Let a spacetime (\ref{metric-Buchdahl}) be a solution with a perfect fluid obeying a barotropic equation of state $p=p(\rho)$ and suppose that it admits a Killing horizon $H(x_{\rm h})=0$.
Then, $p=\rho=0$ holds at $x=x_{\rm h}$ unless $p=-\rho \ne 0$ is satisfied there.
\end{Coro}
{\it Proof:}
The energy-momentum tensor of a perfect fluid is given by Eq.~(\ref{Tab-a}) with $p_1=p_2=:p$, namely
\begin{align}
{T}_{\mu\nu}=&(\rho+p)u_\mu u_\nu +pg_{\mu\nu}.\label{Tab-pf}
\end{align}
By Proposition~\ref{Prop:matter-on-horizon}, an energy-momentum tensor at $x=x_{\rm h}$ can be written as Eq.~(\ref{Tab-1}), which is a mixture of a particular anisotropic fluid and a null dust fluid.
A perfect fluid (\ref{Tab-pf}) is compatible with an anisotropic fluid in Eq.~(\ref{Tab-1}) only if $p=-\rho=p_2$.
Therefore, if a perfect fluid obeying a barotropic equation of state $p=p(\rho)$, $p=-\rho \ne 0$ or $p=\rho=0$ holds at $x=x_{\rm h}$.
\qed
\bigskip

By Corollary~\ref{Coro:p-fluid}, if a perfect fluid obeys a linear equation of state $p=(\gamma-1)\rho$ with $\gamma\ne 0$, $p=\rho=0$ must hold on a Killing horizon.
Let us see two such examples of perfect-fluid solutions in four dimensions given as
\begin{align}
&\D s^2=-H(x)\D t^2+\frac{\D x^2}{H(x)}+r(x)^2\D\Omega^2,\label{Buchdhal-4}\\
&u^\mu=(H(x)^{-1/2},0,0,0),\label{comoving}
\end{align} 
where $\D\Omega^2:=\D\theta^2+\sin^2\theta\D\varphi^2$.
It should be emphasized that the above solution in the comoving coordinates (\ref{comoving}) is valid only in the region with $H(x)> 0$.
If the spacetime (\ref{Buchdhal-4}) admits a regular Killing horizon, the spacetime can be extended beyond there.
In a trapped region defined by $H(x)<0$, the corresponding matter field is no more a perfect fluid but an anisotropic fluid.
By Proposition~\ref{Prop:matter-on-horizon} and Corollary~\ref{Coro:p-fluid}, the matter field on a Killing horizon is a null dust fluid if $r''(x_{\rm h})\ne 0$.
If $r''(x_{\rm h})=0$ holds, there is no matter field on a Killing horizon.

The first example is a special class of the Whittaker solution~\cite{Whittaker} obeying $p=-\rho/3$.
This solution is given by Eqs.~(\ref{Buchdhal-4}) and (\ref{comoving}) with 
\begin{align}
&H(x)=1+\frac{2\beta m}{\tan(\beta x)},\label{whittaker-H}\\
&r(x)=\left|\beta^{-1}\sin(\beta x)\right|,\label{whittaker}\\
&\rho=-3p=3\beta^2H(x),\label{whittaker-rho}
\end{align} 
where $m$ and $\beta$ are constants and the solution reduces to the Schwarzschild vacuum solution in the limit $\beta\to 0$.
As Corollary~\ref{Coro:p-fluid} claims, $p=\rho=0$ certainly holds on a Killing horizon $x=-\beta^{-1}\arctan(2\beta m)(\equiv x_{\rm h})$.
Because of 
\begin{align}
-r^{-1}r''=\beta^2(>0),
\end{align} 
there exists a null dust fluid at $x=x_{\rm h}$ with positive energy density by Eq.~(\ref{mu-horizon}).

Another example is the Semiz solution obeying $p=-\rho/5$~\cite{Semiz:2020lxj}.
This solution can be described by Eqs.~(\ref{Buchdhal-4}) and (\ref{comoving}) with 
\begin{align}
&H(x)=\biggl(1-\frac{2m}{x}\biggl)\biggl\{1-\frac{\lambda x^2}{3}\biggl(1-\frac{2m}{x}\biggl)^3\biggl\}^{-1},\label{semiz-H} \\
&r(x)=\left|x\biggl\{1-\frac{\lambda x^2}{3}\biggl(1-\frac{2m}{x}\biggl)^3\biggl\}\right|,\label{SemizI}\\
&\rho=-5p=5\lambda H(x)^2,\label{general2}
\end{align} 
where $\lambda$ and $m$ are constants and the solution reduces to the Schwarzschild vacuum solution for $\lambda=0$.
Again, $p=\rho=0$ certainly holds on the Killing horizon $x=2m(\equiv x_{\rm h})$.
In this Semiz solution, we obtain $r''(x_{\rm h})=0$ and therefore a matter field is absent at $x=x_{\rm h}$ due to Eq.~(\ref{mu-horizon}).

In the Buchdahl coordinates (\ref{Buchdhal-4}), asymptotic behaviors of the type-I matter field (\ref{bounce-rho})--(\ref{bounce-p2}) (with $n=4$ and $k=1$) toward a Killing horizon $x\to x_{\rm h}$ were investigated in~\cite{Bronnikov:2008ia}.
Theorem~1 in~\cite{Bronnikov:2008ia} states that, in the limit $x\to x_{\rm h}$, the type-I matter field obeys (i) $p_1/\rho\to -1$ and $\rho(x)\to \rho_{\rm h}$ with a constant $\rho_{\rm h}$ or (ii) $p_1/\rho\to -1/(1+2N)$ and $\rho(x)\propto (x-x_{\rm h})^N$ with a natural number $N$.
The case (ii) includes the Whittaker solution (\ref{whittaker-H})--(\ref{whittaker-rho}) for $N=1$ and the Semiz solution (\ref{semiz-H})--(\ref{general2}) for $N=2$.

\section{Summary}

In the present paper, we have shown that, contrary to popular belief, a matter field on a Killing horizon defined by $H(x_{\rm h})= 0$ in a ``static'' spacetime (\ref{metric-Buchdahl}) can be of the Hawking-Ellis type II if $r''\ne 0$ holds there.
Even in such a case, inequalities of the standard energy conditions in the region with $H(x)\ne 0$ can be conveniently used on the Killing horizon just by taking the limit $x\to x_{\rm h}$.
As a consequence, $r''(x)> 0$ is a sufficient condition to violate all the standard energy conditions in the spacetime (\ref{metric-Buchdahl}) including Killing horizons.

We have also exposed a generic property of ``static'' perfect-fluid solutions admitting a Killing horizon, which is independent from the asymptotic behavior or energy conditions.
If a perfect fluid obeys a barotropic equation of state $p=p(\rho)$, $p=\rho=0$ holds on the Killing horizon $x=x_{\rm h}$ unless $p=-\rho \ne 0$ is satisfied there.
Then, there exists a null dust fluid at $x=x_{\rm h}$ if and only if $r''(x_{\rm h})\ne 0$ holds.

The present paper has revealed that singular coordinate systems may lead to incorrect conclusions on the properties of Killing horizon.
In the four-dimensional spherically symmetric case, the matter field on a Killing horizon has been erroneously claimed to be of type I in~\cite{Visser:1992qh} based on a singular coordinate system.
Subsequently, the same claim has been made in the most general static~\cite{Medved:2004ih} and stationary spacetimes~\cite{Medved:2004tp} based on singular coordinate systems where the inverse metric diverges on a Killing horizon and these results have been used in a recent paper~\cite{Martin-Moruno:2021niw}.
However, as shown in this paper, we definitely need to adopt regular coordinate systems on a Killing horizon in order to obtain a correct result in these spacetimes.
These tasks are left for future investigations.




\end{document}